\documentclass[a4paper,11pt]{article}
\usepackage{graphicx}				
\usepackage{amsmath}				
\usepackage{amssymb}
\usepackage{subfigure}				
\usepackage{url}				
\usepackage[numbers,sort&compress]{natbib}	
\usepackage[american]{babel}		
\begin{document}
\selectlanguage{american}
\title{Comparison of Various Methods for the Calculation of the Distance Potential Field}
\author{Tobias Kretz, Cornelia B{\"o}nisch, and Peter Vortisch  \\ \\
PTV AG\\
Stumpfstra{\ss}e 1\\ D-76131 Karlsruhe, Germany\\ \\
{\tt \{Tobias.Kretz,Cornelia.Boenisch,Peter.Vortisch\}@PTV.De}\\
}
\maketitle
\begin{abstract}
The distance from a given position toward one or more destinations, exits, and way points is a more or less important input variable in most models of pedestrian dynamics. Except for the special case when there are no obstacles in a concave scenario -- i.e. each position is visible from any other -- the calculation of these distances is a non-trivial task. This isn't that big a problem, as long as the model only demands the distances to be stored in a \emph{Static Floor Field} also called \emph{Potential Field} \cite{Khatib1986,Latombe1991,Burstedde2001,Schadschneider2001b,Kirchner2002b,Kirchner2002,Nishinari2004,Kretz2006f}, which never changes throughout the whole simulation. In this case a pre-calculation once before the simulation starts is sufficient. But if one wants to allow changes of the geometry during a simulation run -- imagine doors or the blocking of a corridor due to some hazard -- in the \emph{Distance Potential Field}, calculation time matters strongly. This contribution gives an overview over existing and new exact and approximate methods to calculate a potential field, analytical investigations for their exactness, and tests of their computation speed. The advantages and drawbacks of the methods are discussed.
\\
\noindent
\end{abstract}

\section{Introduction}
\subsection{General Introduction}
The will to move through space is the will to reach some kind of destination. On a smaller time scale this is the will to reduce the distance toward some kind of destination. Therefore it appears to be natural to use the gradient of the distance toward a destination in some kind of measure as primary input and impetus for motion in the simulation of the movement of pedestrians.

It is assumed that it is always sufficient to have a discrete distance potential field, either because the model itself is formulated in a discrete manner, or because some finite -- yet arbitrarily large -- exactness of the distance potential field is sufficient. Although potential fields are by no means confined to rectangular grids \cite{Meister2007}, only rectangular grids are investigated.

\subsection{Robots and Pedestrians}
Note that the potential discussed in this contribution has the meaning and is used in the way as in the simulation of pedestrian dynamics \cite{Burstedde2001,Schadschneider2001b,Kirchner2002b,Kirchner2002,Nishinari2004,Kretz2006f,Schadschneider2009} and not robotics \cite{Khatib1986,Latombe1991}. The difference is that in robotics it is usually assumed that an \emph{autonomous robot} knows about his destination coordinate but has no knowledge of the position of obstacles except for those which it ``sees''. For pedestrians on the contrary it is typically assumed that they have at least some knowledge on the whole path, the positions of obstacles and the detours they have to walk compared to linear distance, even if they actually only see a small fraction of the whole path. This has consequences for the calculation and use of the potential.

\begin{figure}[htbp]
  \centering
  \includegraphics*[width=0.55\textwidth]{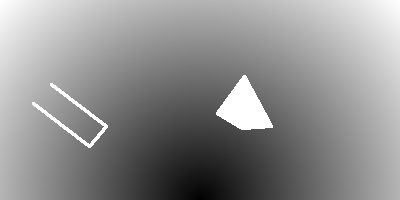} 
 	\caption{Example for $U_{\vec{x}_d}(\vec{x})$ from equation (\ref{eq:U}).}
	\label{fig:khatibpotential}
\end{figure}

In robotics the \emph{artificial potential} at position $\vec{x}$ from a destination at $\vec{x}_d$ was originally \cite{Khatib1986} calculated as
\begin{eqnarray}
U_{artificial}(\vec{x}) & = & U_{\vec{x}_d}(\vec{x}) + U_{obstacles}(\vec{x})\\
U_{\vec{x}_d}(\vec{x}) & = & \frac{1}{2} k_d (\vec{x} - \vec{x}_d)^2 \label{eq:U}
\end{eqnarray}
leading to a potential as shown in figure \ref{fig:khatibpotential}. In such a potential local minima can occur and a robot has to be equipped with the ability to realize it is in a minimum and how to get out of it.

\begin{figure}[htbp]
  \centering
  \includegraphics*[width=0.55\textwidth]{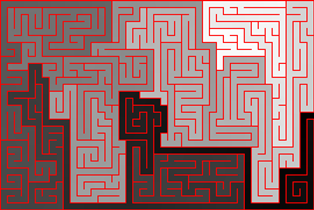} 
 	\caption{For pedestrians the \emph{static floorfield} or \emph{distance potential} is calculated differently to account for the global knowledge pedestrians have in most situations. Mazes pose an exception inasmuch as the essence of their existence is that it is hard, if not impossible, to acquire perfect global knowledge.}
	\label{fig:maze}
\end{figure}

Much different from this is the basic assumption in many models of pedestrian motion that pedestrians have a good global knowledge of the exact Euclidean distances and know about the shortest path between their current position and their destination under consideration of all obstacles. This assumption can be unrealistic for very complex geometries like huge mazes (compare figure \ref{fig:maze}), but for many situations it comes close to reality.

Note: the notion \emph{Distance Potential Field} was chosen to clarify that it is nothing but a direct look-up table for the distance, whereas the \emph{Static Floorfield} is inversely proportional to the distance and therefore -- although edited in a trivial way -- a quantity that's derived from distance to destination.

\section{Methods for the Calculation of a Distance Potential Field}
\subsubsection*{Short Mathematical Parenthesis: Vector Norms} 
In two dimensions the so-called \emph{p-norms} are defined as
\begin{equation}
||\vec{x}||_p := \left(|\Delta x|^p + |\Delta y|^p\right)^{\frac{1}{p}}\text{, with }p \in \mathbb{R}_{+}
\end{equation}
of these $p = 2$ is the well known \emph{Euclidean Metric} (Pythagorean Theorem). So, for $p = 2$ the norm has the everyday life meaning of the word ``distance". At first two other well known norms, which are relevant here, will be discussed: the ones with $p = 1$ and the limit $p \rightarrow \infty$. In principle the problem can be stated this way: only the metrics for $p = 1$ and $p \rightarrow \infty$ can be calculated by a flood fill, what one wants, however, is the metric with $p=2$ for which no equally fast and simple calculation method exists.

In this paper the following notation is used: the distance $D$ from a certain point to the exit on the shortest path for a method $X$ is called $D^X$. It is composed of a sum of distances $d_i^X$ of (straight) line segments of the visibility graph \cite{deBerg1997}, i.e. they connect sequently the exit, corners of obstacles, and the coordinate under consideration. Note that for the flood fill methods the visibility graph does not have to be known explicitly for the calculations. 

\subsection{Flood Fill Methods}
In \emph{Flood Fill} (sometimes also called \emph{Wavefront}) methods the distance is calculated by sequently moving cell to closest neighbor cell and by that summing up the distances.

\subsubsection{Manhattan Metric}
For $p = 1$ one has the famous \emph{Manhattan Metric} -- also called \emph{Taxicab Metric} or \emph{Manhattan Distance}. It was introduced by Hermann Minkowski -- but note that the name ``Minkowski Metric" is reserved for the elementary metric of special relativity. Just as a taxi driver in Manhattan needs to sum up the number of Streets and the number of Avenues which he has to cross during the drive to get an estimation of the distance, the distance $D^M$ according to the Manhattan metric simply is the sum of the absolute values of the differences in x- and y-coordinate. Flood Fill therefore only acts within the \emph{von Neumann Neighborhood}, i.e. grid cells need to be connected via a common edge (see figure \ref{fig:Manhattan}).
\begin{eqnarray}
\Delta x&=&\sum_i{|\delta x_i|} \text{ and } \Delta y=\sum_i{|\delta y_i|}\\
D^M&=&\sum_i{d_i^M} = \Delta x + \Delta y
\end{eqnarray}

\subsubsection{Chessboard Metric}
\begin{figure}[htbp]
 \centering
  \subfigure[Manhattan metric]{
    \label{fig:Manhattan}
    \includegraphics*[width=0.36\textwidth]{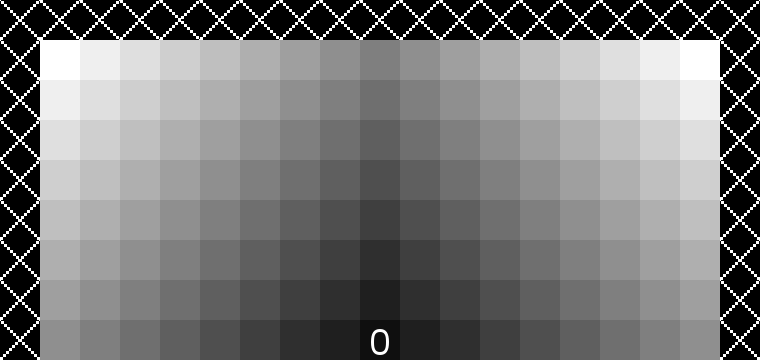} 
  }
  \subfigure[Chessboard metric]{
    \label{fig:Chessboard}
    \includegraphics*[width=0.36\textwidth]{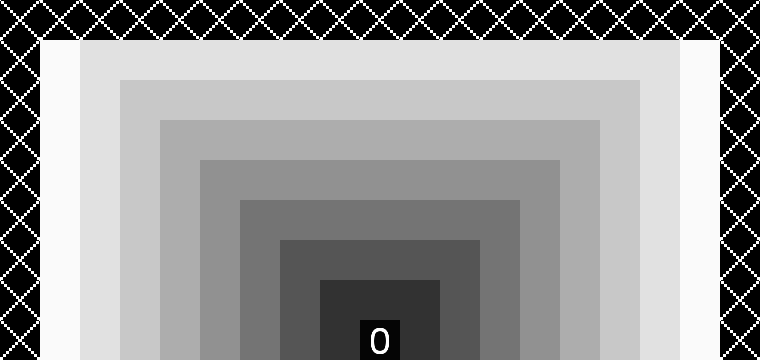} 
  }
	\caption{Manhattan and Chessboard metric.}
	\label{fig:MC}
\end{figure}

For the limit $p \rightarrow \infty$ one arrives at the \emph{Maximum Norm} - also called \emph{Chebyshev Distance} $D^C$ or \emph{Chessboard Metric}. It was introduced by Pafnuty Chebyshev and got its alternative names from the way the king in chess is allowed to move, respectively the fact that the distance measured with this norm is the maximum of the differences in x- or y-coordinate. In other words: flood fill acts within the \emph{Moore Neighborhood} and the added value is the same (typically 1) whether grid cells are connected via an edge or a corner (see figure \ref{fig:Chessboard}).
\begin{eqnarray}
d_i^C&=&\max\left(|\delta x_i|,|\delta y_i|\right)\\
D^C&=&\sum_i{d_i^C}
\end{eqnarray}

The advantage of Manhattan and Chessboard metric is the fact that the typically very fast flood fill procedure can be applied. For $2 < p < \infty$ it is in general not possible to calculate the correct distance (i.e. the Euclidean distance) between two arbitrary grid cells with a flood fill with local rules. The major drawback of course is that aside from Manhattan, Mannheim, or chess there are only few occasions where one of the two norms gives an exact result. However, there are possibilities to stick with the flood fill method and gain exactness. Some of these will be discussed in the following.

\subsubsection{Variant 1: Combination of Manhattan and Chessboard} 
\begin{figure}[hbtp]
  \centering
  \subfigure[``Minimum metric"]{
    \label{fig:Min_norm}
    \includegraphics*[width=0.36\textwidth]{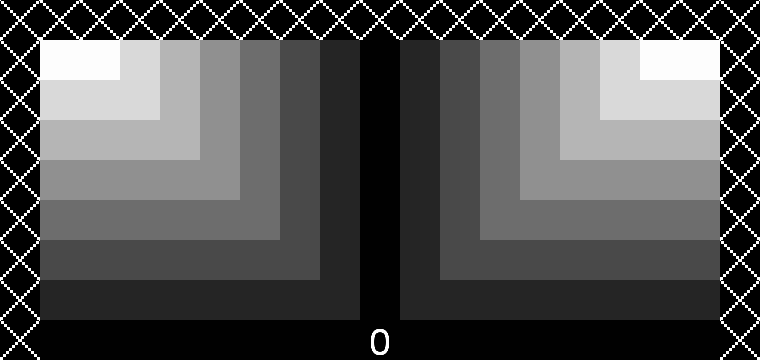} 
  }
  \subfigure[Variant 2]{
    \label{fig:V2}
    \includegraphics*[width=0.36\textwidth]{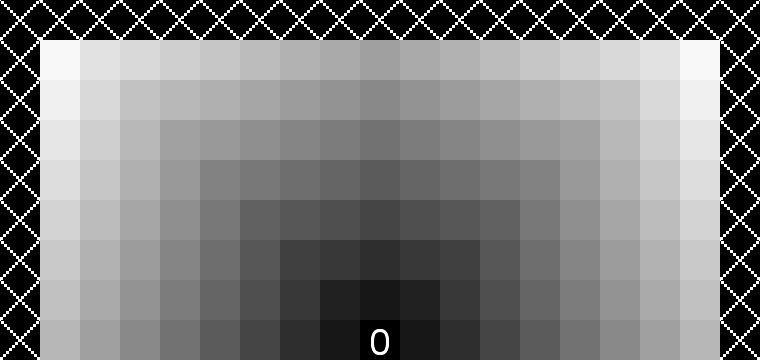} 
  }
	\caption{``Minimum metric" and metric for variant 2}
	\label{V1V2}
\end{figure}

There is a method which uses flood fill and which is exact in the Euclidean sense for all positions which are visible from the destination. Since Manhattan metric gives $d^M_{i=1} = |\delta x_{i=1}| + |\delta y_{i=1}|$ and Chessboard metric results in $d^C_{i=1} = \max \left(|\delta x_{i=1}|, |\delta y_{i=1}|\right)$, one can simply calculate $d^m_{i=1} = d^M_{i=1} - d^C_{i=1} = \min \left(|\delta x_{i=1}|, |\delta y_{i=1}|\right)$. This ``Minimum Norm" (see figure \ref{fig:Min_norm}) is of no use in itself, but $\left(d^m_{i=1}\right)^2 + \left(d^C_{i=1}\right)^2$ must equal $\left(\delta x_{i=1}\right)^2 + \left(\delta y_{i=1}\right)^2$, as one of the two $\delta$s as trivially as necessarily needs to be the maximum and the other one the minimum. And $\left(\delta x_{i=1}\right)^2 + \left(\delta y_{i=1}\right)^2$ is the square of the exact Euclidean distance ($p=2$) between the exit and a coordinate which is directly visible from the exit. If one generalizes this for line segments which are not directly visible from the exit ($i>1$) one makes an error, since one can only calculate the square root of the sum of squared line segment (Euclidean) distances, while the exact result would be the sum of the square roots of squared line segment (Euclidean) distances:

\begin{equation}
\sqrt{\sum_i{(d_i^E)^2}}\neq \sum_i\sqrt{(d_i^E)^2}
\end{equation}

Regardless of this problem, equations (\ref{eq:V1a}) -- (\ref{eq:V1b}) can be motivated by the initial observation that the calculation of distances is exact ``to the point of the next corner". This is achieved at the expense of making two flood fills -- the final value of $D^{V_1}$ can not be calculated by a single flood fill -- and having to calculate a square root for each cell in the end.
\begin{eqnarray}
d^m_i &=& d^M_i - d^C_i = \min\left(|\delta x_i|,|\delta y_i|\right) \label{eq:V1a}\\
D^m &=& \sum_i{d_i^m} = D^M - D^C \\
\left(D^{V_1}\right)^2 &=& \left(D^C\right)^2 + \left(D^m\right)^2  \label{eq:V1b}
\end{eqnarray}
\vspace{-12pt}
\vspace{-12pt}

\subsubsection{Variant 2: $\sqrt{2}$ over Corners} 
Another simple modification is a Chessboard flood fill where flooding across corners adds a $\sqrt{2}$ instead of $1$. If one does so, not only distances parallel to the discretization axis will be exact, but also distances deviating by exactly $45^\circ$ from that. A version of such a modification, which additionally includes a smoothing mechanism, is introduced in \cite{Kluepfel2003}. 
Since the number of diagonal steps is $\min(\delta x_i, \delta y_i)$ and the number of horizontal or vertical steps is $\max(\delta x_i, \delta y_i)-\min(\delta x_i, \delta y_i)$ one can write
\begin{eqnarray}
d^{V_2}_i &=& \sqrt{2} d^m_i + \left(d^C_i - d^m_i\right)\\
D^{V_2} &=& \sum_i{d_i^{V_2}} = \left(\sqrt{2}-1\right) D^m + D^C
\end{eqnarray}

\subsubsection{Variant 3: Larger Neighborhoods}
A modification which gains computation speed on the cost of exactness is to increase the neighborhood further, so cells with a distance of two or three or even more are included. As one only has to continue the calculation with the border cells as new center cells, the recursion depth or stack size is reduced, but one runs the risk of overlooking small obstacles. This method is not investigated further here.

\subsection{Dijkstra's Algorithm on a Visibility Graph}
\begin{figure}[htbp]
  \centering
  \includegraphics*[width=0.55\textwidth]{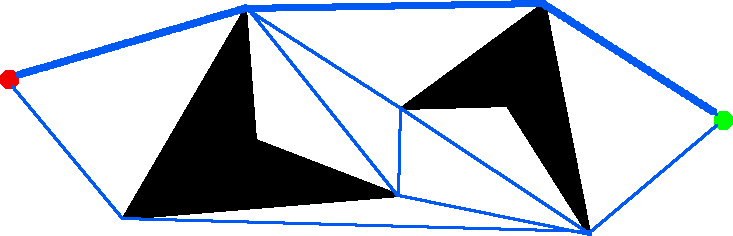} 
 	\caption{Visibility graph of a simple geometry.}
	\label{fig:visgraph}
\end{figure}

Another method is to try to find a subset of grid cells which form a \emph{Visibility Graph} \cite{deBerg1997,Nishinari2004,Kretz2007} (compare figure \ref{fig:visgraph}). These grid cells are all grid cells which are necessary as navigation cell for at least one arbitrary grid cell if a pedestrian wants to move from that arbitrary grid cell around the obstacles to the destination. Two nodes of the visibility graph are connected if and only if they are mutually visible. Once one has created such a visibility graph, one can calculate the distance toward the destination for all grid cells which are part of the visibility graph using \emph{Dijkstra's Algorithm} \cite{Dijkstra1959}. After having done that one can calculate the distance from all other grid cells toward the destination by making use of the visibility graph and the distance information now contained within it. Note that strictly spoken the method used to measure the computation time is the one from \cite{Kretz2007} and not from \cite{deBerg1997,Nishinari2004}, where the latter one is probably more efficient.

\subsection{Ray Casting}
Another very intuitive method is an iterated \emph{Ray Casting}. In detail, the following algorithm is applied:
\begin{enumerate}
\item	Calculate the distance toward the destination for all grid cells which are visible from the destination.
\item	If possible, find that grid cell $X_0$, which is the closest to the destination of all those grid cells which have been assigned a distance, but which have at least one neighbor which is neither an obstacle, nor has been assigned a distance toward the destination, or has been assigned a distance, but a distance which is too large.
\item	Calculate the distance toward the destination of all cells which are visible from $X_0$. If there are cells which are visible to $X_0$ and which have already been assigned a distance toward the destination, this distance is only overwritten, if the newly calculated distance is smaller.
\item	Repeat steps 2. and 3. until step 2. does not find any grid cell anymore.
\end{enumerate}

It is important for the calculation time how the check for visibility is done. Concerning the ray tracing part one can use the \emph{Bresenham Line Drawing Algorithm} \cite{Bresenham1965}. But it is important that this algorithm is not used to draw a line (cast a ray) each time one wants to calculate the visibility of a grid cell. It is better to draw a rectangle around the whole scenario (``border") and cast rays from the cell in focus to each of those border cells. All cells before the first obstacle are then marked as visible, all behind as ``not visible". Because of the discreteness, it might occur that the ray casting toward neighbored border cells gives different results for the visibility of some cell (when a cell is part of multiple casts). In that case, the cell needs to be marked as visible, because otherwise in scenarios with narrow spaces ``blind spots" can appear, of which the algorithm would claim that they are not accessible at all. With this strategy the number of lines, one has to draw, only grows as the border size instead of as the area.

\subsection{Other Methods of Error Reduction}
One has to distinguish the {\em distance error} in the distance potential field from the {\em speed error} in a model. If the pedestrians move on a discrete lattice as well, the lattice structure can lead to errors in the speed, just as it leads to errors in the distance. The speed errors can be compensated for by making pedestrians suspend certain moves depending on the ratio of corner versus edge steps they did in recent moves. Such strategies are proposed for example in \cite{Kluepfel2003,Schultz2006}.

\section{Analytical Considerations}
\subsection{Errors for Manhattan and Chessboard Metric}
The maximal errors of the two simple metrics are both: well known and trivial to calculate, nevertheless for the sake of completeness, they are given in the following.
The absolute error compared to the Euclidean distance $d^E_i$ of a single straight line element of some path using the Manhattan metric is
\begin{eqnarray}
e^M_i&=&d^M_i - d^E_i\\
     &=&d^E_i(|\cos{\varphi_i}|+|\sin{\varphi_i}|-1)\\
\text{with } d^E_i &=& \sqrt{|\delta x_i|^2 + |\delta y_i|^2} 
\end{eqnarray}
with $\varphi_i$ being the angle between the connecting line of the two points and the x-axis.
This simply sums up to a total error of
\begin{equation}
E^M=\sum_i{d^E_i(|\cos{\varphi_i}|+|\sin{\varphi_i}|-1)}
\end{equation}
which always lies between the boundaries 
\begin{eqnarray}
0&\leq& E^M \leq (\sqrt{2}-1)D^E\\
\text{with } D^E&=&\sum_i{d^E_i} \text{\hspace{24 pt}(exact total Euclidean distance)}
\end{eqnarray}
The maximal error arises from diagonal motion.\par
The corresponding values for the Chessboard metric are
\begin{eqnarray}
e^C_i&=&d^C_i - d^E_i\\
     &=&d^E_i(\max{(|\cos{\varphi_i}|,|\sin{\varphi_i}|)}-1)\\
E^C&=&\sum_i{d^E_i(\max{(|\cos{\varphi_i}|,|\sin{\varphi_i}|)}-1)}
\end{eqnarray}
with the latter one always satisfying the relation 
\begin{equation}
(\sqrt{0.5}-1)D^E\leq E^C \leq 0
\end{equation}
Here as well the extreme value is reached for diagonal motion.

\subsection{Error for Variant 1 (Combination)}
To calculate the (extremal) errors, one has to deal with in variant 1, is a bit more complicated than it is for the two basic metrics above. The error is
\begin{eqnarray}
E^{V_1} &=& D^{V_1} - D^E\\
&=& \sqrt{\Big(\sum_i{\max{\left(|\delta x_i|,|\delta y_i|\right)}}\Big)^2+\Big(\sum_i{\min{\left(|\delta x_i|,|\delta y_i|\right)}}\Big)^2} \label{eq:Variant1a}\\
&& - \sum_i{\sqrt{|\delta x_i|^2+|\delta y_i|^2}} \nonumber
\end{eqnarray}
Remember that
\begin{equation}
|\delta x_i| = d^E_i|\cos\left(\varphi_i\right)| \text{ and }|\delta y_i| = d^E_i|\sin\left(\varphi_i\right)|
\end{equation}
and use as abbreviation
\begin{eqnarray}
M_i &=& \max{\left(|\cos\left(\varphi_i\right)|,|\sin\left(\varphi_i\right)|\right)} \label{eq:max}\\
m_i &=& \min{\left(|\cos\left(\varphi_i\right)|,|\sin\left(\varphi_i\right)|\right)} \label{eq:min}
\end{eqnarray}
in equation (\ref{eq:Variant1a}).
\begin{eqnarray}
E^{V_1}&=& \sqrt{\sum_i{\sum_{j}{d^E_id^E_j\left(M_iM_j+m_im_j\right)}}} - \sum_i{d^E_i}\\
       &=& \sqrt{\sum_i{\left(d^E_i\right)^2} + \sum_i{\sum_{j\neq i}{d^E_id^E_j\left(M_iM_j+m_im_j\right)}}} - \sum_i{d^E_i}\\
       &=& \sqrt{\Big(\sum_i{d^E_i}\Big)^2 + \sum_i{\sum_{j\neq i}{d^E_id^E_j\left(M_iM_j+m_im_j-1\right)}}} - \sum_i{d^E_i}\\
       &=& D^E\left(\sqrt{1 + \frac{\sum_i{\sum_{j\neq i}{d^E_id^E_j\left(M_iM_j+m_im_j-1\right)}}}{\left(\sum_i{d^E_i}\right)^2}} - 1\right)
\end{eqnarray}
With the relations
\begin{eqnarray}
0\leq &m_i \leq \sqrt{0.5} \leq M_i& \leq 1\\
\sqrt{0.5}\leq &m_im_j + M_iM_j& \leq 1
\end{eqnarray}
one gets for the error
\begin{equation}
D^E\left(\sqrt{1 + \left(\sqrt{0.5}-1\right)\frac{\sum_i{\sum_{j\neq i}{d^E_id^E_j}}}{\left(\sum_i{d^E_i}\right)^2}} - 1\right) \leq E^{V_1} \leq 0
\end{equation}
which is extremal for direction changes from horizontal/vertical to diagonal or vice versa. No error at all arises, when the new direction can be generated from the old by a reflection at one of the axis or diagonals.
The left part will take the most extreme value, if all of the $d^E_i$ are equal. With $N$ as the number of single straight line elements one gets
\begin{equation}
D^E\left(\sqrt{1 + \left(\sqrt{0.5}-1\right)\frac{N-1}{N}} - 1\right) \leq E^{V_1} \leq 0
\end{equation}
This confirms the initial observation that there is no error for $N=1$. However, the error in this variant depends on the number of line elements and therefore the number of obstacles in -- respectively the complexity of -- the scenario. In the limit $N \rightarrow \infty$ the error can in the worst case (denoted by the hat) be 
\begin{equation}
\hat{E}^{V_1} = D^E\left(0.5^{0.25} - 1\right) \approx -0.159D^E
\end{equation}

\subsection{Error for Variant 2 ($\sqrt{2}$ over Corners)}
For variant 2 the error is
\begin{eqnarray}
E^{V_2} &=& \left(\sqrt{2}D^m + D^C - D^m\right) - D^E\\ 
				&=& \left(\left(\sqrt{2}-1\right)D^m + D^C\right) - D^E\\
        &=& \sum_i{\left(\left(\sqrt{2}-1\right) \min(|\delta x_i|,|\delta y_i|) + \max(|\delta x_i|,|\delta y_i|) - d_i^E\right)}
\end{eqnarray}
with equations (\ref{eq:max}) and (\ref{eq:min}) this is
\begin{eqnarray}
E^{V_2}&=& \sum_i{d_i^E\left(\left(\sqrt{2}-1\right) m_i + M_i - 1\right)}\\ 
			 &=& \sum_i{d_i^E\left(\left(\sqrt{2}-1\right) m_i + \sqrt{1-m_i^2} - 1\right)}
\end{eqnarray}
bearing in mind that $0\leq m_i\leq \sqrt{0.5}$ one finds the maximum of each summand at
\begin{equation}
\hat{m}_i = \frac{\sqrt{2-\sqrt{2}}}{2}
\end{equation}
which corresponds to an angle of exactly $\hat{\phi} = \pi/8 = 22.5^\circ$ and any corresponding angle in the other seven octants.
The maximum error then is
\begin{eqnarray}
\hat{E}^{V_2}&=& \left(\sqrt{4-2\sqrt{2}}-1\right)D^E\\
             &\approx& 0.082D^E
\end{eqnarray}
which is slightly better than the maximum error $\hat{E}_2^{V_1} \approx 0.099$ for variant 1 for $N=2$
With this method there can never be a negative error. If one wants the error to vanish on the average of all angles, one can add the values
\begin{eqnarray}
s_{hv} &=& \frac{\hat{\varphi}}{\alpha}=\frac{\pi}{8\left(\sqrt{2}-1\right)} \approx 0.948\\
s_{d} &=& \sqrt{2} \frac{\hat{\varphi}}{\alpha}=\sqrt{2} \frac{\pi}{8\left(\sqrt{2}-1\right)} \approx 1.341
\end{eqnarray}
when flooding horizontally or vertically respectively diagonally. I.e. there is a global factor of $\approx 0.948$ multiplied to any distance, as $s_d / s_{hv} = \sqrt{2}$.
In this case the directions $\varphi_{1,2}$ (of the first octant) with exact measurements are at
\begin{equation}
\sin{\varphi_{1,2}} = \frac{4-2\sqrt{2}}{\pi} \pm \sqrt{\frac{1}{4-2\sqrt{2}} - \frac{8}{\pi^2}}
\end{equation}
which is approximately $\varphi_1 \approx 9.55^\circ$ and $\varphi_2 \approx 35.45^\circ$. In the range between these two angles distances are measured too large, outside of this range too small.
\section{Computation Times}
\subsection{Geometries}
\begin{figure}
  \center
	\includegraphics[width=0.3\textwidth]{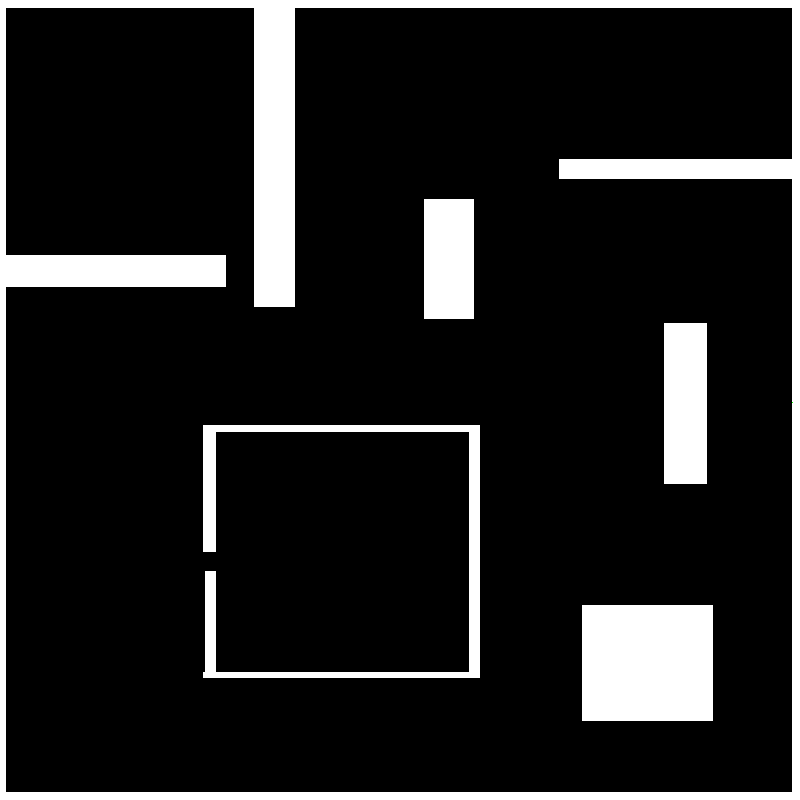} \hspace{10pt}
	\includegraphics[width=0.3\textwidth]{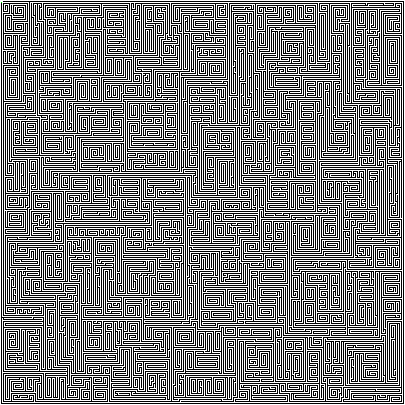} \hspace{10pt}
	\includegraphics[width=0.3\textwidth]{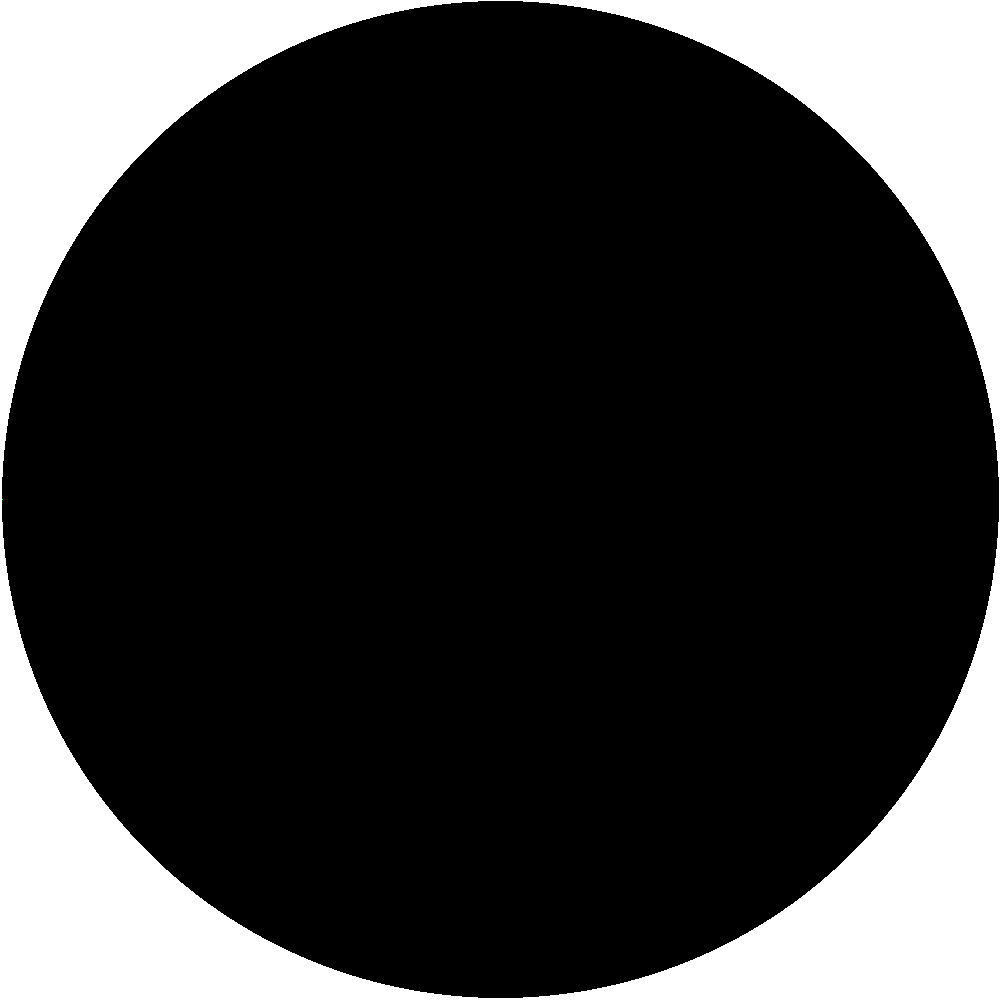}\\ \vspace{10pt}
	\includegraphics[width=0.3\textwidth]{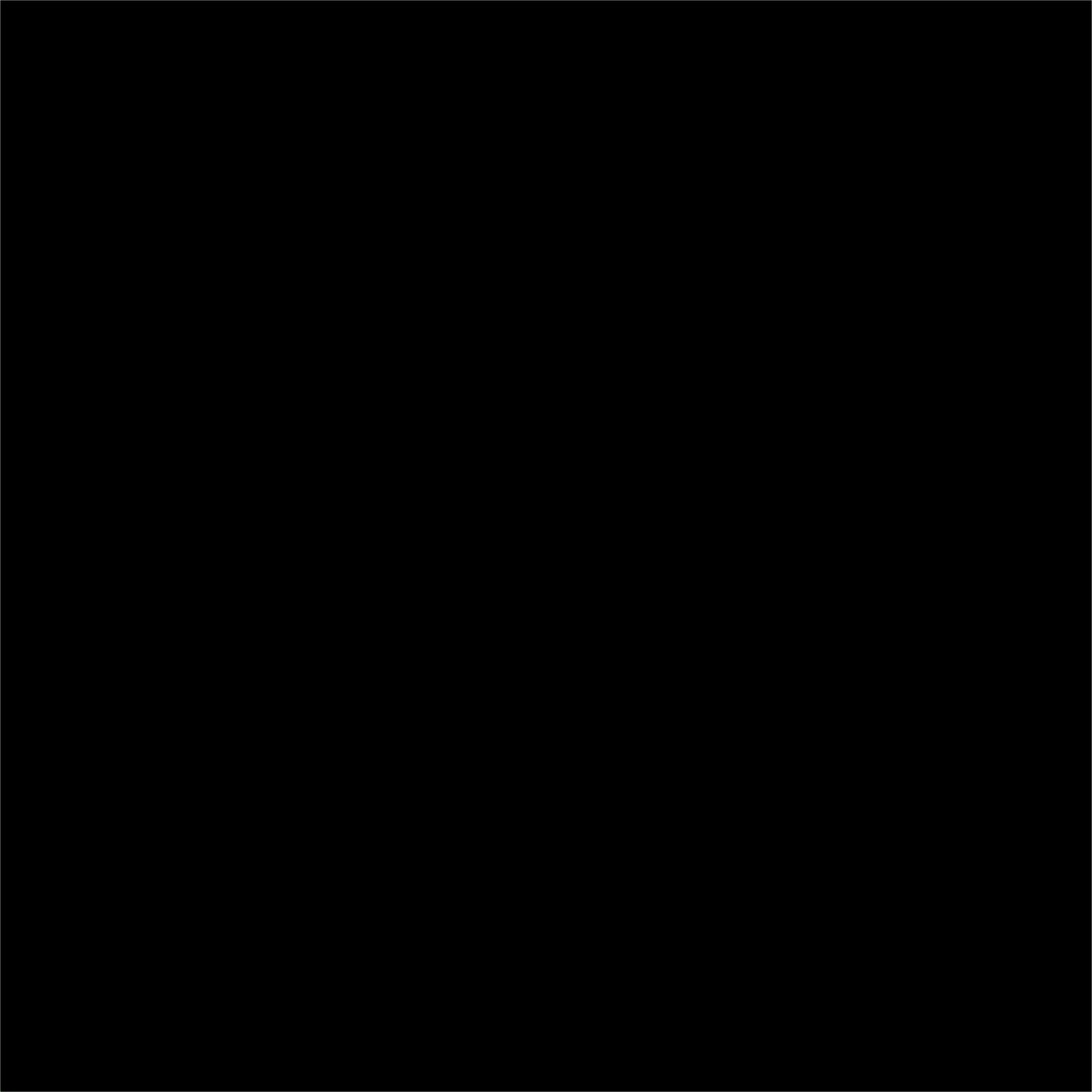} \hspace{10pt}
	\includegraphics[width=0.3\textwidth]{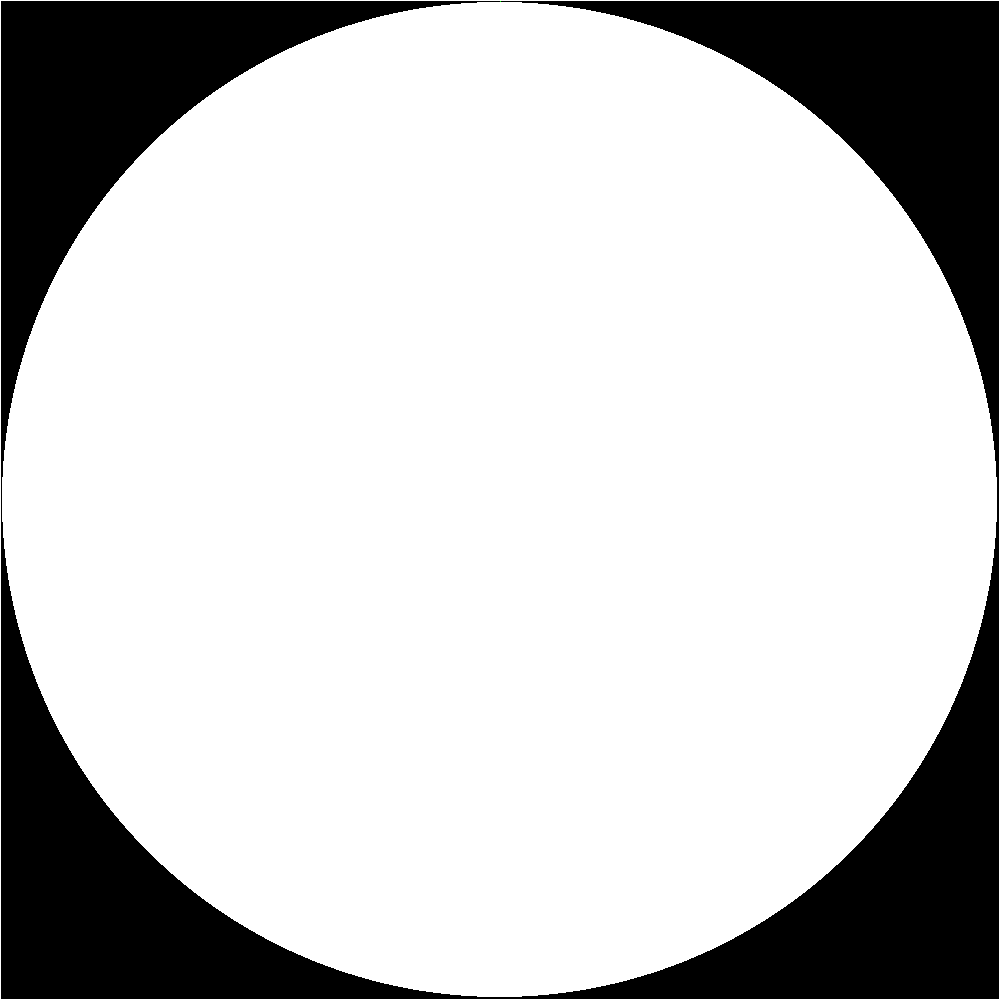} \hspace{10pt}
	\includegraphics[width=0.3\textwidth]{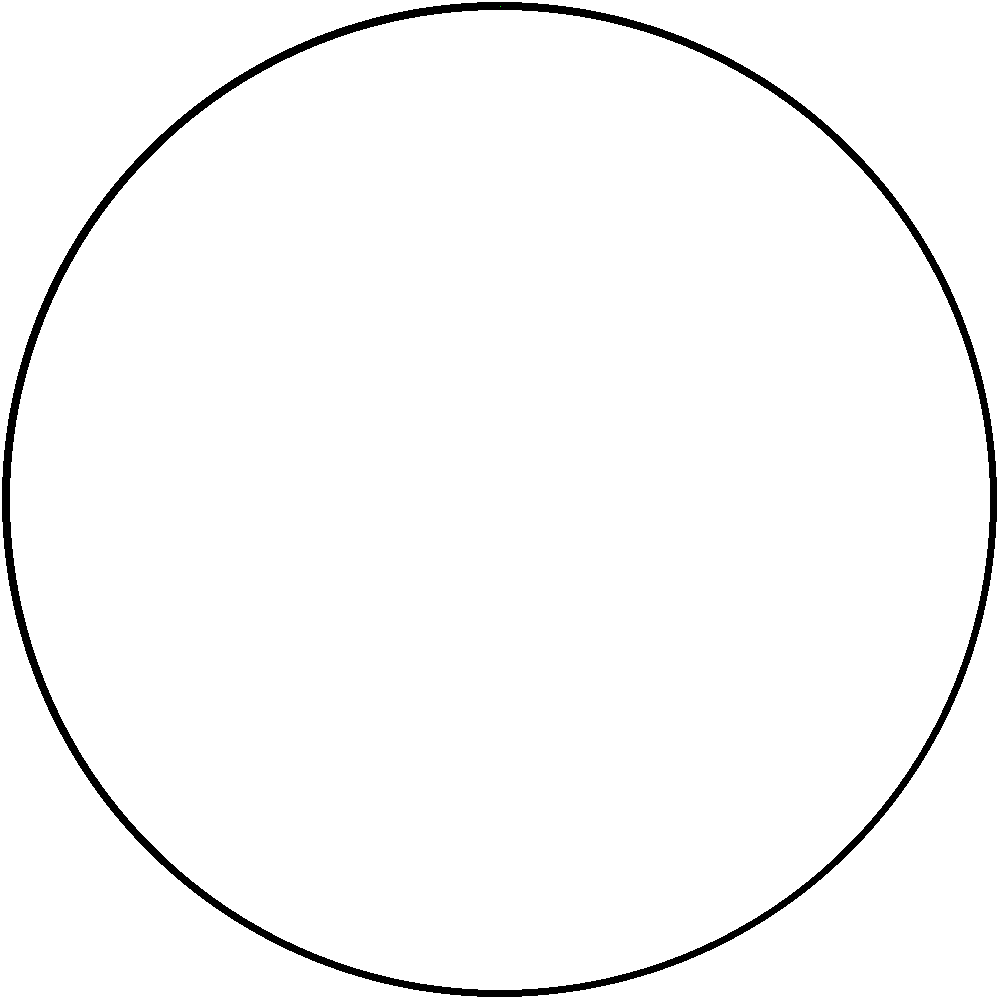}
	\caption{Test geometries. Walkable areas are colored black, walls white. Note that the scenarios were largely different in size and just scaled differently to fit the page (compare description in the text).}
	\label{fig:geometries}
\end{figure}

The test geometries were -- as shown in figure \ref{fig:geometries} -- a ``typical" room, a maze with 200 x 200 grid cells, a circle shaped room with a diameter of 996 grid cells , a square shaped room with a side length of 3998 grid cells, a room with a large column in the middle, and a ring, the latter two with the same size as the circle.

\subsection{Results}
The following tables give the results for computation time (standard PC) and the largest distance from the exit that was found by the method. Among the flood fill based methods variant 2 is different from all other flood fill methods regarding the fact that taking a square root for each grid cell is a necessary part of the calculation. For square roots and other elements of the calculation the calculation time not only depends on the algorithm but also on the details of the implementation. 

In the visibility graph method, the visibilities were calculated via the area method. This led to the comparatively large calculation times. Though most of the time they were smaller than those of the ray casting method with visibility calculation by the area method. The only exception was the circle scenario. The reason for this is that the node generating method presented in \cite{Kretz2007} generates many unnecessary nodes, as the bending of the circle's border was so small that many local neighborhoods might also have been part of a convex corner, although the inner border of a circle is entirely concave. If the circle had been given as geometric object this could have been avoided.

\begin{table}
\center
\begin{tabular}{|l|r|r||l|r|r|} \hline
\bf{Geometry: Typical}		&Calc.      &Maximal &\bf{Maze} &Calc.      &Maximal\\ 
Method                  &Time [s]		&Distance&     &Time [s]		&Distance\\ \hline 
Manhattan               &0.04       &1246.00 &     &0.00       &7507.00\\
Chessboard     		      &0.06       & 960.00 &     &0.01       &7198.00\\
Variant 1 (Comb.)     	&0.12       &1001.70 &     &0.02       &7204.63\\
Variant 2 ($\sqrt{2}$)  &0.08       &1091.35 &     &0.01       &7325.97\\
Ray Casting (edge) 		  &1.18       &1052.85 &     &72.72      &7361.20\\
Visibility Graph				&11.00  		&1052.16 &	   &386.2	     &7325.97\\
Ray Casting (area)      &51.00      &1053.72 &     &1156       &7507.00\\ \hline\hline
\bf{Geometry: Circle}   &						&				& \bf{Square} &					 &				\\ \hline
Manhattan               &0.06       &1202.00&        &2.44       &7994.00\\
Chessboard    		      &0.10       & 996.00&        &2.69       &3997.00\\
Variant 1 (Comb.)	      &0.18       & 996.24&        &5.52       &5652.61\\
Variant 2 ($\sqrt{2}$)  &0.10       &1037.52&        &2.68       &5652.55\\
Ray Casting (edge) 		  &0.34       & 996.24&  		   &4.29       &5652.61\\
Visibility Graph				&100.80     & 996.24& 			 &628.00     &5652.61\\
Ray Casting (area)      &19.00      & 996.24&        &1606.00    &5652.61\\ \hline\hline
\bf{Geometry: Column}    &						&				& \bf{Ring}  &						&\\ \hline
Manhattan               &0.02       &1993.00&       &0.01       &1971.00\\
Chessboard    		      &0.03       &1408.00&       &0.01       &1386.00\\
Variant 1 (Comb.)     	&0.06       &1524.69&       &0.02       &1504.40\\
Variant 2 ($\sqrt{2}$)  &0.03       &1650.30&       &0.01       &1628.30\\
Ray Casting (edge) 		  &8.89       &1565.08&  		  &7.36       &1542.24\\
Visibility Graph				&47.80      &1564.94& 		  &7.00       &1542.09\\
Ray Casting (area)      &163.00     &1565.26&       &22.00      &1542.39\\ \hline
\end{tabular}\vspace{12pt}

\caption{Computation time and maximal distance for the geometries of figure \ref{fig:geometries}. }
\label{tab:results}
\end{table}

The ray casting method which uses the border method for visibility calculations was pleasingly fast, although never as fast as the Manhattan flood fill, and probably not yet fast enough to be applied each time step for many destinations.

\section{Conclusions}
In this contribution various methods for the calculation of distances in an obstacle filled space were investigated for their deviation from the true Euclidean distance and the time consumption for their calculation. Starting with the two well-known metrics -- Manhattan and Chessboard -- variants of such flood fill methods were reviewed respectively introduced. Following that a visibility graph and a ray casting method were discussed. It was found that compared to the standard metrics it is possible to significantly reduce the error while sticking to the flood fill method by making subtle changes to or using combinations of the standard metrics. The errors that remain balance with the fact that their calculation is significantly quicker than the calculation of the two other methods which are -- in principle -- error-free.

\section*{Acknowledgments:}
For valuable discussion and hints we thank U. Hanebeck, S. Hengst, A. Pohlmann, and L. Spannehl.
%
\bibliographystyle{unsrt}
\bibliography{Kretz}

\end{document}